\documentclass[12pt,dvips]{article}
\usepackage{graphicx}
\renewcommand{\descriptionlabel}[1]%
	     {\hspace{\labelsep}\textsf{#1}}

\def\slashii#1{\setbox0=\hbox{$#1$}             
   \dimen0=\wd0                                 
   \setbox1=\hbox{\sl/} \dimen1=\wd1            
   \ifdim\dimen0>\dimen1                        
      \rlap{\hbox to \dimen0{\hfil\sl/\hfil}}   
      #1                                        
   \else                                        
      \rlap{\hbox to \dimen1{\hfil$#1$\hfil}}   
      \hbox{\sl/}                               
   \fi}               

\def\slashiii#1{\setbox0=\hbox{$#1$}#1\hskip-\wd0\hbox to\wd0{\hss\sl/\/\hss}}

\def\slashi#1{\rlap{\sl/}#1}

\newcommand{\be}{\begin{eqnarray}}
\newcommand{\ee}{\end{eqnarray}}

\catcode`\@=11
\def\lsim{\mathrel{\mathpalette\@versim<}}
\def\gsim{\mathrel{\mathpalette\@versim>}}
\def\@versim#1#2{\vcenter{\offinterlineskip
\ialign{$\m@th#1\hfil##\hfil$\crcr#2\crcr\sim\crcr } }}
\catcode`\@=12
\begin{document}


\thispagestyle{empty}
\mbox{}
\vspace{2cm}
\vskip1truecm
\centerline{\Large\bf  Neutrino Physics}
\vskip1truecm
\centerline{\sc $^{\;a}$Paul Langacker\footnote{E-mail: pgl@electroweak.hep.upenn.edu}, 
  $^{\;b}$Jens Erler\footnote{E-mail: erler@fisica.unam.mx}, and  $^{\;b}$Eduardo Peinado\footnote{E-mail: epeinado@nirvana.fisica.uson.mx} }
\vspace{6pt}
 \centerline{\it $^a$Department of Physics, University of Pennsylvania, Philadelphia, PA 19104, USA}

\centerline{\it $^b$Instituto de F\'\i sica, Universidad Nacional Aut\'onoma
                de M\'exico, 01000 M\'exico D.F., M\'exico}
                
 \vspace{6pt}

\bigskip

\centerline{{\bf Abstract}}
The theoretical and experimental bases of neutrino mass and mixing are reviewed. A brief chronological evolution of the weak interactions, the electroweak Standard Model, and neutrinos is presented. Dirac and Majorana mass terms are explained as well as models such as the seesaw mechanism. Schemes for two, three and four neutrino mixings are presented. 

\vspace{3cm}

\centerline{\sl $11^{th}$ Mexican School}
\centerline{\sl of Particles and Fields}
\centerline{\sl August 2004}
\vskip1truecm



\thispagestyle{empty}

\mbox{}

\newpage

\pagenumbering{roman}

\tableofcontents

\newpage

\pagenumbering{arabic}

\setcounter{page}{1}

\title{Neutrino Physics}
\section{The Fermi Theory}
The history of the weak interactions dates back to the discovery of radioactivity by Becquerel in 1896~\cite{weakints}.
In particular, $\beta$ decay, in which a nucleus emits an electron and increases its charge,
apparently violated the conservation of energy (as well as momentum and, as we now understand,
angular momentum). In 1931 Pauli postulated that a massless, chargeless, essentially
non-interacting particle that he
named the ``neutron''  (later renamed the neutrino by Fermi) was also emitted in the process
and carried off the missing energy. Pauli's hypothesis was verified around 1953
when the electron-type neutrino (actually the anti-neutrino
$\bar{\nu}_e$) produced in a reactor was observed directly by its rescattering by Reines
and Cowen. The second (muon-type) neutrino, $\nu_\mu$, associated with the
$\mu$ in its interactions, was detected by its rescattering to produce a muon
 in 1962 by Lederman, Schwartz, and
Steinberger at Brookhaven. The  third charged lepton, the $\tau$, was discovered at SLAC in
1975. There was ample indirect evidence from the weak interactions of the $\tau$ that 
an  associated neutrino, $\nu_\tau$,
must exist, but it  was not observed directly until 2000 at Fermilab.

In 1934, Enrico Fermi developed a theory of $\beta$ decay.
The Fermi theory is loosely like QED, but of zero range (non-renormalizable) and non-diagonal (charged currents).
\begin{figure}[h]
\begin{center}
\includegraphics[height=4cm]{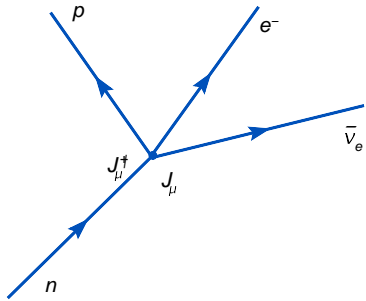}
\caption{Diagram for $\beta$-decay in the Fermi theory}
\label{betad}
\end{center}
\end{figure}
The Hamiltonian  (see Figure~\ref{betad}) is
\begin{equation}
H \sim \frac{G_F}{\sqrt{2}} J^{\dag}_{\mu}J^{\mu},
\end{equation}
where $J_{\mu}$ is the charged current,
\[
\begin{array}{llclc}
J^{\dag}_{\mu} &\sim& \bar{p} \gamma _{\mu} n& + &\bar{\nu}_{e} \gamma _{\mu} e^{-} \\
&&[n \longrightarrow p &,& e^{-}\longrightarrow \nu_{e}]
\end{array},
\]
and $G_{F}$ is the Fermi constant, with the modern value $G_{F}=1.16637(1)\times10^{-5}$~GeV$^{-2}$. The Fermi theory was later rewritten 
in terms of quarks, and  modified to include: $\mu^{-}$ and $\tau^{-}$ decays;  strangeness changing transitions (Cabibbo); 
the heavy quarks $c$, $b$, and $t$; quark mixing and CP violation (the Cabibbo-Kobayashi-Maskawa matrix); vector ($V$) and 
axial vector ($A$) currents (i.e, $V\sim \gamma_\mu$ is replaced by $V-A\sim \gamma_\mu(1-\gamma^{5})$), which implies parity violation
(Lee-Yang; Wu; Feynman-Gell-Mann); and $\nu$  mass and mixing.

The Fermi theory gives an excellent description (at tree-level) of such  processes as:
\begin{itemize}
\item Nuclear/neutron $\beta-$decay: $n \longrightarrow p e^{-}\bar{\nu}_{e}$.
\item $\mu^{-}$, $\tau^{-}$ decays: $\mu^{-} \longrightarrow e^{-}\bar{\nu}_{e}\nu_{\mu}$; \ $\tau^{-} \longrightarrow \mu^{-}\bar{\nu}_{\mu}\nu_{\tau},\  \nu_{\tau} \pi^-, \cdots $.
\item $\pi$, $K$ decays: $\pi^{+} \longrightarrow \mu^{+}\nu_{\mu}, \ \pi^{0}e^{+}\nu_{e}$; $K^{+} \longrightarrow \mu^{+}\nu_{\mu}, \  \pi^0 e^+ \nu_e, \ \pi^+\pi^0, \cdots$.
\item Hyperon decays: $ \Lambda \longrightarrow p \pi^{-}$,\ $\Sigma^{-} \longrightarrow n \pi^{-}$, \ $\Sigma^{+} \longrightarrow  \Lambda e^{+} \nu_{e}, \cdots $.
\item Heavy quarks decays: 
$b \longrightarrow c \mu^{-}\bar{\nu}_{\mu},\ c\pi^{-}; t\longrightarrow b \mu^+\nu_\mu, \cdots$.
\item Neutrino scattering: $\nu_{\mu}e^{-} \longrightarrow \mu^{-} \nu_{e}$,  $ \underbrace{\nu_{\mu} n \longrightarrow \mu^{-} p}_{\mbox{elastic}}$, $ \underbrace{\nu_{\mu} n \longrightarrow \mu^{-} X}_{\mbox{deep inelastic}}, \cdots$.
\end{itemize}

\begin{figure}[h]
\begin{center}
\includegraphics[height=4cm]{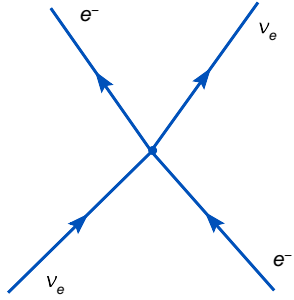}
\caption{$\nu_e-e^{-}$ scattering}
\label{nue1}
\end{center}
\end{figure}
\begin{figure}[h]
\begin{center}
\includegraphics[height=5cm]{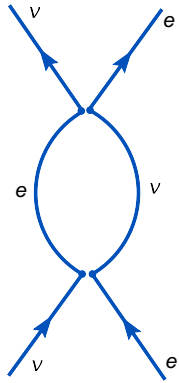}
\caption{Second order interaction in the Fermi theory}
\label{f2o}
\end{center}
\end{figure}

However, the Fermi theory violates unitarity at high energy.
Consider the process $\nu_{e}e^{-} \longrightarrow \nu_{e}e^{-}$ (see Figure~\ref{nue1}), which is described by the effective Lagrangian,
\begin{equation}
{\cal L}=-\frac{G_{F}}{\sqrt{2}}J^{\dag}_{\mu}J^{\mu}.
\label{efla}
\end{equation}
The amplitude has only the $L=0$ partial wave, and the cross section, $\sigma$, is proportional to $\frac{G_{F}^2 s}{\pi}$, where 
${\sqrt{s}}=E_{cm}$ is the total energy in the center of mass reference frame. However, for a pure S-wave process unitarity requires 
$\sigma < \frac{16 \pi}{s}$. Thus, for energies, 
$$\frac{1}{2}E_{cm} \gsim \sqrt{\frac{\pi}{G_{F}}} \sim 500\mbox{ GeV},$$ the theoretical $\sigma$
would violate unitarity. The non-unitarity of Born-approximation amplitudes is usually restored by higher order terms. However, 
the Fermi theory involves divergent integrals in second order, such as,
$$ \int d^{4} k \frac{\slashi k + m_{e}}{k^{2}-m_{e}^{2}} \frac{\slashi k}{k^{2}}, $$
which corresponds to the process in Figure~\ref{f2o}. 
Moreover, it is non-renormalizable due to the dimension of the coupling $[G_{F}]=(\mbox{mass})^{-2}$.

\section{Intermediate Vector Bosons (IVB) and {\boldmath $SU(2)_W\times U(1)_Y$}} \label{ivbsec}
\begin{figure}[h]
\begin{center}
\includegraphics[height=2.7cm]{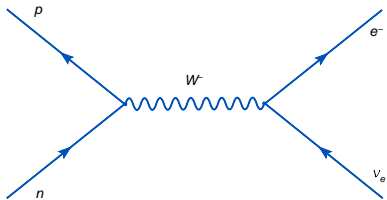}
\caption{The intermediate vector boson}
\label{IVB}
\end{center}
\end{figure}

The intermediate vector boson $W^\pm$ was postulated by Yukawa in 1935 in the
same paper as the meson theory, and reintroduced by Schwinger in 1957. 
It is a massive charged particle which mediates the weak
interaction (in analogy with the photon  in QED) as in Figure~\ref{IVB}.

Assuming that the IVB is massive, in the low energy regime, $M_{W} \gg |Q|$, we have $\frac{G_{F}}{\sqrt{2}}\sim\frac{g^{2}}{8 M_{W}^{2}}$,
where (in modern normalization) the coupling of the $W^{\pm}$ to the current is $\frac{g}{2\sqrt{2}}$. The amplitudes for processes like 
$\nu_{e} e^{-}\longrightarrow \nu_{e} e^{-}$ are now better behaved at high energy because they are no longer pure S-wave. 
However, unitarity violation occurs for $E_{cm}/2 \gsim 500$~GeV for processes involving external $W^\pm$, such as  
$e^{+} e^{-}\longrightarrow W^{+}W^{-}$, because of the growth of the longitudinal polarization vector 
$\epsilon_{\mu} \sim \frac{k_{\mu}}{M_{W}}$. Even though the coupling $g$ is dimensionless, the theory is still nonrenormalizable.

The divergent parts of the diagrams can be canceled if one also introduces a massive neutral boson $W^{0}$, with appropriate $W^{0}W^{+}W^{-}$ and $e^{+}e^{-}W^{0}$
vertices. Requiring such cancellations in all processes leads to 
the relation $[J^{-},J^{+}]=J^{0}$ for the charged and neutral currents, i.e., the couplings are those of an $SU(2)$ gauge model. However, the model is not realistic because it does not
incorporate QED. In 1961 Glashow~\cite{glashow} developed a gauge model with the symmetry $SU(2)\times U(1)$, in which there was the $ W^{\pm}$ and a massive neutral  $Z$ boson, as well the photon, $\gamma$, but there was no mechanism for generating the masses of the $W^{\pm}$ and $Z$ bosons and the fermions
(bare masses would destroy renormalizability). In 1967 Weinberg~\cite{weinberg} and 
Salam~\cite{salam} in 1968 independently applied the idea of Higgs~\cite{higgs} on how massless gauge particles can acquire mass through spontaneous symmetry breaking to the $SU(2)_W\times U(1)_Y$ model\footnote{The subscripts $W$ and $Y$ refer to {\em weak} and {\em hypercharge}, respectively.}. In 1971 't Hooft 
and Veltman~\cite{thooft} proved the renormalizability of spontaneously broken gauge theories. 

The $SU(2)_W\times U(1)_Y$ model worked well for leptons and predicted the existence
of weak  neutral current (WNC) transitions such as $\nu_e \longrightarrow \nu_e$ and 
$e^- \longrightarrow e^-$ mediated by the new $Z$ boson.
However, the extension to hadrons (quarks)
was problematic: the observed Cabibbo-type mixing between the $s$ and $d$ quarks in the
charged current implied flavor changing neutral current (FCNC)
transitions $d\leftrightarrow s$ mediated by the $Z$
because of the different  transformations of the $d$ (doublet) and $s$ (singlet) under
$SU(2)_W$. These were excluded by the non-observation of certain rare $K$ decays and
because of the size of the $K^0-\bar{K}^0$ mixing (for which there was also a too large
contribution from a second order charged current diagram).
In 1970, Glashow, Iliopoulos and Maiani~\cite{Glashow:1970gm} proposed a new quark (the charm, $c$, quark) as the $SU(2)$ partner of the $s$, avoiding the FCNC and $K^0-\bar{K}^0$ problems.
The $c$ quark 
 was discovered in 1974 through the observation of the $J/\Psi$ meson. The (strangeness-conserving) WNC was discovered in 1973 by the Gargamelle collaboration at CERN~\cite{Hasert:1973ff} and by HPW at Fermilab~\cite{HPW}, and was subsequently studied
 in great detail experimentally~\cite{fivephases}, 
 as were the weak interactions of heavy quarks, especially the $b$.
 The $W$ and $Z$ were observed directly with the predicted masses in 1983 by the
 UA1~\cite{ua1} and UA2~\cite{ua2}
  experiments at CERN, and the $SU(2)_W\times U(1)_Y$ theory was 
 probed at the radiative correction level in high precision experiments at LEP (CERN) and
 SLC (SLAC) from 1989 until $\sim 2000$~\cite{precisionstatus}. The existence of non-zero neutrino mass and
 mixing was firmly established by the Super-Kamiokande collaboration
 in 1998~\cite{skatm}, and intensively studied subsequently.
 
\section{Aspects of the {\boldmath $SU(2)_W\times U(1)_Y$} Theory}
It is convenient to define the  left (right) chiral projections of a fermion field $\Psi$
by  $\Psi_{L(R)} \equiv \frac{1}{2}(1 \mp \gamma^{5})\Psi$. $\Psi_{L(R)}$ coincide with
states of negative (positive) helicity for a massless fermion, while chirality and 
helicity differ in amplitudes by terms of $O(m/E)$ for relativistic fermions of mass $m$ and energy
$E$.
In the electroweak Standard Model (SM)~\cite{weakints,Langacker:2005ar} the left ($L$)
and right ($R$) chiral fermions are assigned to transform respectively as  
doublets and  singlets of $SU(2)_W$ in order to incorporate the observed $V-A$ structure
of the weak charged current. 
The left-chiral quark and lepton doublets are
\begin{equation}
\begin{array}{lcccl}
\\
q^{0}_{mL}=
\left(
\begin{array}{c}
u^{0}_{m}\\
d^{0}_{m}
\end{array}
\right)_{L}
&\mbox{     }&\mbox{  and   }&\mbox{     }&
l^{0}_{mL}=
\left(
\begin{array}{c}
\nu^{0}_{m}\\
e^{-0}_{m}
\end{array}
\right)_{L}
\end{array},
\end{equation}
where the $0$ superscript refers to weak eigenstates, i.e., fields associated to weak transitions, $m=1,\ldots F$ is the family index, and $F$ is the number of families.
The right-chiral singlets are $u_{mR}^{0}$, $d_{mR}^{0}$, $e_{mR}^{- 0}$, ($\nu_{mR}^{0}$). 
The weak eigenstates will in general be related to the mass eigenstate fields by
unitary transformations.  The quark color indices $\alpha=r,g,b$ have been suppressed (e.g., $u_{m\alpha R}^{0}$) to simplify the notation.
The fermionic part of the weak Lagrangian is
{\small
\begin{equation}
 {\cal L}_{F} =  \displaystyle \sum_{m=1}^{F}\left( \bar{q}^{0}_{mL}i \slashiii D q^{0}_{mL} + \bar{l}^{0}_{mL}i \slashiii D l^{0}_{mL}
+ \bar{u}^{0}_{mR}i \slashiii D u^{0}_{mR} + \bar{d}^{0}_{mR}i \slashiii D d^{0}_{mR} + \bar{e}^{0}_{mR}i \slashiii D e^{0}_{mR} \right). \label{lferm}
\end{equation}}
The assignment of the chiral $L$ and $R$ fields to different representations leads to parity violation, which is maximal for $SU(2)_W$, and also implies that bare mass terms are forbidden by the gauge symmetry. The gauge covariant derivatives are given by
\begin{equation}
\begin{array}{lll}
D_{\mu}q^{0}_{mL} &=&  \left( \partial_{\mu}+\frac{ig}{2}\tau^{i}W^{i}_{\mu}+\frac{ig'}{6}B_{\mu}\right)q^{0}_{mL}, \\
D_{\mu}l^{0}_{mL} &=&  \left( \partial_{\mu}+\frac{ig}{2}\tau^{i}W^{i}_{\mu}-\frac{ig'}{2}B_{\mu}\right)l^{0}_{mL}, \\
D_{\mu}u^{0}_{mR} &=&  \left( \partial_{\mu}+i\frac{2}{3}g'B_{\mu}\right)u^{0}_{mR}, \\
D_{\mu}d^{0}_{mR} &=&  \left( \partial_{\mu}-i\frac{g'}{3}B_{\mu}\right)d^{0}_{mR}, \\
D_{\mu}e^{0}_{mR} &=&  \left( \partial_{\mu}-ig'B_{\mu}\right)e^{0}_{mR},
\end{array} \label{covariant}
\end{equation}
where $g$ and $g'$ are respectively the $SU(2)_W$ and $U(1)_Y$ gauge couplings, while $W^i, i=1..3,$ and $B$ are the (massless) gauge 
bosons. The weak hypercharge ($U(1)_Y$) assignments are determined by $Y=Q-T^3$. The right-handed fields do not couple to the $SU(2)_W$ 
terms. One can easily extend~(\ref{lferm}) and (\ref{covariant}) by the addition of SM-singlet  right-chiral neutrino fields
$\nu_{mR}^{0}$, with $ D_{\mu}\nu_{mR}^{0}= \partial_{\mu}\nu_{mR}^{0}$.

The Higgs mechanism may be used to give masses to the gauge bosons and chiral fermions.
In particular, let us introduce
 a complex doublet of scalar fields,
$$\phi=\left(\begin{array}{c} \phi^{+} \\ \phi^{0}\end{array}\right),$$
and assume that the potential for $\phi$ is such that
 the neutral component acquires a vacuum expectation value (VEV)\footnote{For a single Higgs doublet one can always perform a gauge rotation so that only the neutral component acquires an expectation value.}. Then (in unitary gauge)
$$\phi \rightarrow \frac{1}{\sqrt{2}} \left(\begin{array}{c} 0 \\  \nu +H \end{array}\right),$$
where $\nu=\sqrt{2}\langle \phi^{o}\rangle =246$~GeV, and $H$ is the physical Higgs scalar. Three of the original four $SU(2)_W\times U(1)_Y$ gauge bosons will become massive and one, the photon,  remains massless. 

If we rewrite the Lagrangian in the new vacuum, the scalar kinetic energy terms take the form
\begin{equation}
\begin{array}{lcl}
\left(D_{\mu} \phi\right)^{\dag}D^{\mu} \phi&\sim& \frac{1}{2}(\begin{array}{cc}0 &\nu \end{array})
\left[\frac{g}{2}\tau^{i}W^{i}_{\mu}+\frac{g'}{2}B_{\mu}\right]^{2}\left(
\begin{array}{c}
0 \\ 
\nu 
\end{array}
\right) \\ &\rightarrow&
M_{W}^{2}W^{+\mu}W^{-}_{\mu}+\frac{M_{Z}^{2}}{2}Z^{\mu}Z_{\mu},
\end{array}
\end{equation}
where the gauge interaction and kinetic energy terms of the physical $H$ particle have been omitted. The mass eigenstate gauge bosons are
\begin{equation}
\begin{array}{lcl}
W^{\pm}&=&\frac{1}{\sqrt{2}}\left(W^{1} \mp iW^{2}\right),\\
Z&=&-\sin \theta_{W}B+\cos \theta_{W} W^{3},\\
A&=&\cos \theta_{W}B+\sin \theta_{W}W^{3},
\end{array}
\end{equation}
where the weak angle is defined by $\tan \theta_{W} \equiv g'/g$.
The masses of the gauge bosons are predicted to be
\[
\begin{array}{lcccr}
M_{W}=\frac{g\nu}{2}, & &M_{Z}=\sqrt{g^{2}+g'^{2}} \frac{\nu}{2},& & M_{A}=0.
\end{array}
\]
$W^\pm$  are the IVB of the charged current, 
 $A=\gamma$ corresponds to the photon, and the $Z$ is a massive neutral boson predicted by the theory, which mediates
 the new neutral current interaction. The Goldstone scalars transform into the longitudinal components of the $W^{\pm}$ and $Z$.

\begin{figure}[h]
\begin{center}
\includegraphics[height=2.7cm]{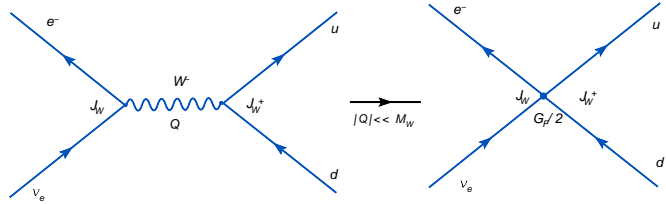}
\caption{Diagram for $\nu_e d \longrightarrow e^-u$ and its low energy limit.}
\label{invb}
\end{center}
\end{figure}

A typical weak process is shown in  the Figure~\ref{invb}. The propagator of the $W$ boson goes like $\frac{1}{Q^{2}-M^{2}_{W}}$. For $|Q^{2}| \ll M^{2}_{W}$, $\frac{1}{Q^{2}-M^{2}_{W}}\rightarrow  -\frac{1}{M^{2}_{W}}$, which leads to the effective zero-range four-Fermi interaction  in Eq.~(\ref{efla}), with 
$\frac{G_{F}}{\sqrt{2}} \sim \frac{g^{2}}{8M^{2}_{W}}$ as in the IVB theory.
The Fermi constant is determined from the lifetime of the muon, $\tau^{-1}_\mu\sim\frac{G_{F}^{2}m^{5}_{\mu}}{192\pi^{3}}$, yielding the electroweak scale
\begin{equation}
\nu=2M_{W}/g \simeq (\sqrt{2}G_{F})^{-1/2} \simeq 246\mbox{ GeV}.
\end{equation}
 Comparing the $A$ coupling to QED, one finds $g=\frac{e}{\sin \theta_{W}}$, where $e$ is the electric charge of the positron.

Similarly, the Yukawa couplings of the Higgs doublet to the fermions introduce fermion mass matrices when $\phi$ is replaced
by its vacuum expectation value.

\subsection{The Weak Charged Current}
\begin{figure}[b]
\begin{center}
\includegraphics[height=3cm]{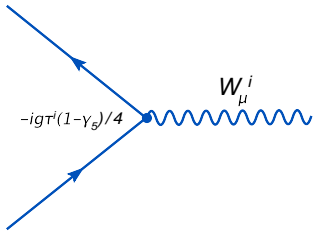}
\caption{$W$-fermion interaction}
\label{current1}
\end{center}
\end{figure}

The major quantitative tests of the SM involve gauge interactions and the properties of the gauge bosons. The charged current interactions
of the Fermi theory are incorporated into the Standard Model and made renormalizable.
The $W$-fermion interaction (Figure~\ref{current1}) is given by
\begin{equation}
{\cal L} = -\frac{g}{2\sqrt{2}}\left( J^{\mu}_{W}W^{-}_{\mu}+J^{\mu \dag}_{W}W^{+}_{\mu}\right),
\end{equation}
where the weak charge-raising current is
\small
\begin{equation}
\begin{array}{lcl}
J^{\mu \dag}_{W}&=& \displaystyle\sum_{m=1}^{F}\left[
\bar{\nu}^{0}_{m}\gamma^{\mu}(1-\gamma^{5})e^{0}_{m}+\bar{u}^{0}_{m}\gamma^{\mu}(1-\gamma^{5})d^{0}_{m}\right]\\
\\
&=&\left(\begin{array}{ccc}\bar{\nu}_{e}&\bar{\nu}_{\mu}&\bar{\nu}_{\tau}\end{array}\right)\gamma^{\mu}(1-\gamma^{5})
\left(
\begin{array}{c}
e^{-}\\
\mu^{-}\\
\tau^{-}
\end{array}
\right)
+
\left(\begin{array}{ccc}\bar{u}&\bar{c}&\bar{t}\end{array}\right)\gamma^{\mu}(1-\gamma^{5})V_{CKM}
\left(
\begin{array}{c}
d\\
s\\
b
\end{array}
\right),
\end{array}
\end{equation}
\normalsize
and where we have specialized to $F=3$ families in the second line and introduced the fields $e,\mu,\tau; u,c,t; d,s,b$ of definite mass 
(mass eigenstates). We are ignoring neutrino masses for now and simply define $\nu_{e}$, $\nu_{\mu}$ and $\nu_{\tau}$ as the weak partners 
of the  $e$, $\mu$ and $\tau$. The pure $V-A$ form ensures maximal P and C violation, while CP is conserved except for phases in $V_{CKM}$.
We define the vectors,
\begin{equation}
u_{L}\equiv
\left(
\begin{array}{c}
u_{L}\\
c_{L}\\
t_{L}
\end{array}
\right), \ \ \ \ \ \ 
u_{L}^0\equiv
\left(
\begin{array}{c}
u_{L}^{0}\\
c_{L}^{0}\\
t_{L}^{0}
\end{array}
\right),
\end{equation}
and the unitary transformations from the weak basis to the mass basis,
\begin{equation}
u_{L}^{0}=A^{u}_{L}u_{L},  \ \ \ \ \ \  d_{L}^{0}=A^{d}_{L}d_{L},  \ \ \ \ \ \  e_{L}^{0}=A^{e}_{L}e_{L},
\label{unitarity1}
\end{equation}
with $A^{u \dag}_{L}A^{u}_{L}=A^{u}_{L}A^{u \dag}_{L}=I$, etc.
In terms of these the unitary quark mixing matrix, due to the mismatch of the unitary transformations
in the $u$ and $d$ sectors, is
\begin{equation}
V_{CKM}= A^{u \dag}_{L} A^{d}_{L}=\left(\begin{array}{ccc}
V_{ud} & V_{us} & V_{ub} \\
V_{cd} & V_{cs} & V_{cb} \\
V_{td} & V_{ts} & V_{tb}\end{array}
\right),
\label{vckm}
\end{equation}
called the Cabibbo-Kobayashi-Maskawa (CKM) matrix~\cite{cabibbo,km}. After removing the unobservable $q_{L}$ phases, the $V_{CKM}$ matrix involves three angles and one CP-violating phase. As described previously, the low energy limit of the charged current interaction 
reproduces the Fermi theory (with the improvement that radiative corrections can be calculated because of the
renormalizability of the theory), which successfully describes weak decays and neutrino scattering processes~\cite{precisiontest}. In particular,
extensive studies, especially in $B$ meson decays, have been done to test the unitarity of $V_{CKM}$ as a probe of new physics and to test 
the  origin of CP violation~\cite{ckmfitter,ckmpdg}. However, 
the CP-violating phase in $V_{CKM}$ is not enough to explain baryogenesis (the cosmological asymmetry between
matter and antimatter), and therefore we need additional sources of CP violation\footnote{Possibilities include electroweak baryogenesis~\cite{baryogenesis} in the supersymmetric extension of the SM, especially
if there are additional scalar singlets, or leptogenesis~\cite{leptogenesis} in theories with a heavy Majorana neutrino.}.

The mixing of the third quark family with the first two is small, and it is sometimes a good zeroth-order approximation
to ignore it and consider an effective $F=2$ theory. Then $V_{CKM}$ is replaced by the  Cabibbo matrix
\begin{equation}
V=\left( \begin{array}{cc} \cos \theta_{c} & \sin \theta_{c}\\
-\sin \theta_{c} & \cos \theta_{c} 
\end{array}
\right),
\end{equation}
which depends on only one parameter,  the Cabibbo angle, $\sin \theta_{c}\simeq0.22$. In this case there is no CP violation because all 
the phases can be reabsorbed in the fields.

In addition to reproducing the Fermi theory, the charged current propagator of the SM has been probed at HERA in
$e^\pm p \longleftrightarrow \stackrel{(-)}{\nu}_e X$ at high energy. The theory has also been tested at the  loop level in processes such 
as $K^0-\bar K^0$ and $B-\bar B$ mixing, and in the calculation of (finite) radiative corrections, which are necessary for the consistency 
of such processes as $\beta$ and $\mu$ decay.

\subsection{The Weak Neutral Current}
The weak neutral current was predicted by the $SU(2)_W \times U(1)_Y$ model and the Lagrangian for it is
\begin{equation}
\begin{array}{lcl}
{\cal L} &=& -\frac{\sqrt{g^{2}+g'^{2}}}{2}J^{\mu}_{Z}\left(-\sin \theta_{W} B_{\mu}+ \cos \theta_{W} W^{3}_{\mu} \right) \\
&=&-\frac{g}{2 \cos \theta_{W}}J^{\mu}_{Z}Z_{\mu},
\end{array}
\end{equation}
\begin{figure}[h]
\begin{center}
\includegraphics[height=3cm]{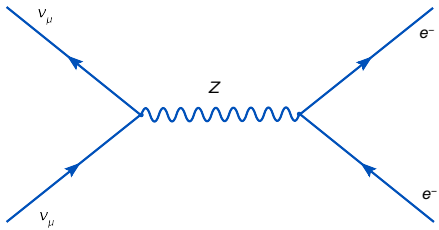}
\caption{The neutral current process $\nu_\mu e^- \longrightarrow \nu_\mu e^-$. 
For $\nu_e e^- \longrightarrow \nu_e e^-$
there is an additional
charged current contribution.}
\label{ncp1}
\end{center}
\end{figure}
where we have used {$\cos \theta_{W}=\frac{g}{\sqrt{g^{2}+g'^{2}}}$}. The neutral current is given by
\begin{equation}
\begin{array}{lcl}
J^{\mu}_{Z} &=& \displaystyle \sum_{m}\left[\bar{u}^{0}_{mL}\gamma^{\mu}u^{0}_{mL}-
\bar{d}^{0}_{mL}\gamma^{\mu}d^{0}_{mL}+\bar{\nu}^{0}_{mL}\gamma^{\mu}\nu^{0}_{mL}-\bar{e}^{0}_{mL}\gamma^{\mu}e^{0}_{mL}
\right] -2\sin^{2}\theta_{W}J^{\mu}_{Q} \\
&=& \displaystyle \sum_{m}\left[\bar{u}_{mL}\gamma^{\mu}u_{mL}-
\bar{d}_{mL}\gamma^{\mu}d_{mL}+\bar{\nu}_{mL}\gamma^{\mu}\nu_{mL}-\bar{e}_{mL}\gamma^{\mu}e_{mL}
\right] -2\sin^{2}\theta_{W}J^{\mu}_{Q},
\end{array}
\end{equation}
where $J^\mu_Q=\Sigma_i q_i \bar{\psi}_i \gamma^\mu\psi_i$   is the electromagnetic current.
 $J^{\mu}_{Z}$ is flavor-diagonal in the Standard Model like the electromagnetic current, i.e., the weak neutral current has the same form in the weak and the mass bases, because the fields which mix have the same assignments under the $SU(2)_W \times U(1)_Y$ gauge group. This
was the reason for introducing the GIM mechanism, discussed in section~\ref{ivbsec}, into the theory. The neutral current has two contributions; one is purely $V-A$, while the other is proportional to the electromagnetic current and is purely vector. Therefore parity is violated but not maximally.

The neutral current interaction of fermions in the low momentum limit is given by
\begin{equation}
{\cal L}^{NC}_{eff}= - \frac{G_{F}}{\sqrt{2}}J^{\mu}_{Z}J_{Z\mu}.
\end{equation}
The coefficients for the charged and  neutral current interactions  are the same because
\begin{equation}
\frac{G_{F}}{\sqrt{2}}=\frac{g^{2}}{8M^{2}_{W}}=\frac{g^{2}+g'^{2}}{8M^{2}_{Z}}.
\end{equation}
Since its discovery, the weak neutral current has been tested and extensively studied in many 
processes~\cite{precisiontest,ewreview,Erler:2004cx,lepewwg,amaldi,costa,llm},
These include pure weak processes such as $\nu e \longrightarrow \nu e$, $\nu N \longrightarrow \nu N$, and $\nu N \longrightarrow \nu X$;  weak-electromagnetic interference, including parity violating asymmetries in processes like
polarized $e^{\uparrow\downarrow} D \longrightarrow e X$, atomic parity violation, and $e^{+}e^{-}$ annihilation 
above and below the $Z$-pole; and high precision (e.g., 0.1\%) {$Z$}-pole reactions at CERN and SLAC.

\section{Neutrino Preliminaries}

Neutrinos are a unique probe of many aspects of physics and astrophysics on scales ranging from $10^{-33}$ to $10^{+28}\ cm$~\cite{nurev}. Decays and scattering processes involving neutrinos have been powerful tests of many aspects of particle physics,
and have been important in establishing the Fermi theory and the SM (at $\sim$ 1\%), searching for
new physics at the TeV scale, as a probe of the structure of hadrons, and as a test of QCD. Small neutrino masses are sensitive to new physics at scales ranging from a TeV up to grand unification and superstring scales. Similarly, neutrinos are important for the physics  of
the Sun, stars, core-collapse supernovae, the origins of cosmic rays, the large scale structure of the universe, big bang nucleosynthesis,
and possibly baryogenesis.

\subsection{Weyl Spinors}

A Weyl two-component spinor is  the minimal fermionic degree of freedom. 
A left-chiral Weyl spinor $\Psi_L$ satisfies
\begin{equation}
P_{L}\Psi_{L}=\Psi_{L}, \ \ \ \ \  P_{R}\Psi_{L}=0,
\end{equation}
where $P_{L(R)}=\frac{1\mp\gamma^{5}}{2}$.
A $\Psi_{L}$ annihilates $L$ particles or creates $R$ antiparticles. Similarly, a right-chiral spinor $\Psi_R$ satisfies 
$P_{R}\Psi_{R}=\Psi_{R}, \ \  P_{L}\Psi_{R}=0$. A Weyl spinor can exist by itself (e.g., the $\nu_{L}$ in the SM), or can be considered 
a projection of a  4-component Dirac spinor $\Psi=\Psi_{L}+\Psi_{R}$ (e.g., $e^- = e^-_L + e^-_R$).
A $\Psi_{L}$ field is related by CP or CPT to a right-chiral antiparticle spinor\footnote{Which is called the particle and which the
antiparticle is a matter of convenience.} $\Psi_{R}^{c}$ ($\Psi_{L}\stackrel{\rm {CP,CPT}}{\longleftrightarrow}\Psi_{R}^{c}$), 
where\footnote{Thus, in our notation the subscript $L$ ($R$) always refers to a left (right) chiral spinor, independent of whether it is 
for a particle or antiparticle.} $\Psi_{R}^{c}$ annihilates $R$ antiparticles or creates $L$ particles and 
$P_{R}\Psi_{R}^{c}=\Psi_{R}^{c}$. Under CP, for example,
\begin{equation}
\Psi_{L}(\vec{x},t) \rightarrow \gamma^{0}\Psi_{R}^{c}(-\vec{x},t).
\end{equation}
$\Psi_{R}^{c}$ is essentially the adjoint of $\Psi_L$:
\begin{equation}
\begin{array}{l}
\Psi_{R}^{c}=C\overline{\Psi_{L}}^{T}=C\gamma_{0}^{T}(\Psi_{L}^{\dag})^{T}, \\
\Psi_{L}=C\overline{\Psi_{R}^c}^{T},
\end{array} \label{adjoint}
\end{equation}
where $T$ represents the transpose and $C$  is the charge conjugation operator, defined by $C\gamma_{\mu}C^{-1}=-\gamma_{\mu}^{T}$. In
the Pauli-Dirac representation
$C=i\gamma^{2}\gamma^{0}$.
The free Weyl field is
\[
\Psi_{L}(x)=\int\frac{d^{3}\vec{p}}{\sqrt{(2\pi)^{3}2E_{p}}}
\left[
a_{L}(\vec{p})u_{L}(\vec{p})e^{-ip \cdot x}+b_{R}^{\dag}(\vec{p})v_{R}(\vec{p})e^{ip \cdot x}
\right] \mbox{  (no sum over spin).}
\]

When two Weyl spinors are present, we can use either ($\Psi_{L}$ and $\Psi_{R}$) or  ($\Psi_{L}$ and $\Psi_{L}^{c}$) to define the theory. In the Standard Model, for example,  baryon ($B$) and lepton ($L$) numbers are conserved perturbatively, and it is convenient
to work with, e.g., 
 $u_{L}$ and $u_{R}$ and similarly for the other fermionic fields. Their CPT partners are $u_{R}^{c} \leftrightarrow  u_{L}$ and $u_{L}^{c} \leftrightarrow  u_{R}$. In some extensions of the SM, such as supersymmetry or grand unification, it is more
 convenient to work with $L$ spinors $\Psi_{L}$ and $\Psi_{L}^{c}$.

 \subsection{Active and Sterile Neutrinos}

Active (a.k.a.~ordinary or doublet) neutrinos are left-handed neutrinos which transform as $SU(2)$ doublets with a charged lepton partner, and which  therefore have normal weak interactions. The
$L$ doublets and their right-handed partners are
\begin{equation}
\begin{array}{lcl}
\left(
\begin{array}{c}
\nu_{e} \\
e^{-}
\end{array}
\right)_{L}
 &\stackrel{\rm  CPT}{\longleftrightarrow}  & 
\left(
\begin{array}{c}
e^{+} \\
\nu_{e}^{c}
\end{array}
\right)_{R}
\end{array}.
\end{equation}

Sterile (a.k.a.~singlet or ``right-handed'') neutrinos, which are present in most extensions of the SM, are $SU(2)$ singlets. They do not interact except by mixing, Yukawa interactions, or beyond the SM (BSM)
interactions. It is convenient to denote the right-chiral spinor as $N_R$ and its conjugate as $N^c_L$.
\begin{equation}
N_{R}\stackrel{\rm  CPT}{\longleftrightarrow} N_{L}^{c}.
\end{equation}
\section{Neutrino Mass Models}
Most extensions of the Standard Model lead to nonzero neutrino mass at some level. 
There are many models of neutrino mass. Here, only a brief survey of the principal classes is given. For more detail see~\cite{massmodels}. 
\subsection{Fermion Masses} 
Mass terms convert a spinor of one chirality into one of the opposite chirality, 
\begin{equation}
 {-\cal L} \sim m(\bar{\Psi}_{2L}\Psi_{1R}+\bar{\Psi}_{1R}\Psi_{2L}).
\end{equation}
An interpretation is that a massless fermion has the same helicity (chirality)  in all frames of reference,
while that of a massive particle depends on the reference frame and therefore can be flipped.

\subsection{Dirac Mass}
\begin{figure}[h]
\begin{center}
\includegraphics[height=1cm]{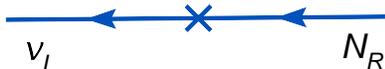}
\caption{Dirac mass term}
\label{diracmass1}
\end{center}
\end{figure}

A Dirac mass (Figure~\ref{diracmass1}) connects two distinct Weyl neutrinos\footnote{These are usually active and sterile.
However, there are variant forms involving two distinct active (Zeldovich-Konopinski-Mahmoud) or two
distinct sterile neutrinos.}
 $\nu_{L}$ and $N_{R}$. It is given by
\begin{equation}
 {-\cal L}_{D} = m_{D}(\bar{\nu}_{L}N_{R}+\bar{N}_{R}\nu_{L})= m_D \bar \nu \nu.
\end{equation}
Thus, $\Psi_{1R}\neq \Psi_{2R}^{c}$ and we have two distinct Weyl spinors.  These may be combined to form a Dirac field $\nu \equiv \nu_{L}+N_{R}$ with four
components: $\nu_{L}$, $\nu_{R}^{c}$, $N_{R}$ and $N_{L}^{c}$. This can be generalized to three or more families.

\begin{figure}[h]
\begin{center}
\includegraphics[height=3cm]{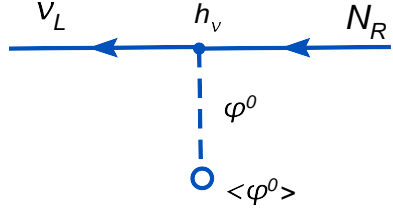}
\caption{Yukawa coupling to the Higgs field}
\label{diracmass2}
\end{center}
\end{figure}

This mass term allows a conserved lepton number $L$, which implies no mixing between $\nu_{L}$ and $N_{L}^{c}$, or between $\nu_{R}^{c}$ and $N_{R}$. However, it violates weak isospin $I_{W}$  by $\Delta I_{W}=\frac{1}{2}$. A Dirac mass can be generated by the Higgs mechanism, as in Figure~\ref{diracmass2}.
The Dirac neutrino masses $m_{D}=h_{\nu}\langle \phi^{0}\rangle =h_{\nu}\frac{\nu}{\sqrt{2}}$ are analogous to the quark and charged lepton masses. The upper bound on the neutrino mass, $m_{\nu}\lsim 1$~eV, implies a Yukawa coupling of the neutrino to the Higgs, $h_{\nu}<10^{-11}$, which is extremely small in comparison with the coupling for the top quark, $h_{t}=O(1)$, or for the electron $h_{e^{-}}\sim 10^{-5}$.

\subsection{Majorana Mass}

\begin{figure}[h]
\begin{center}
\includegraphics[height=1cm]{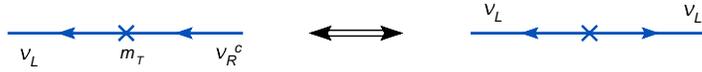}
\caption{A Majorana mass term}
\label{majmass}
\end{center}
\end{figure}

\begin{figure}[h]
\begin{center}
\includegraphics[height=2.7cm]{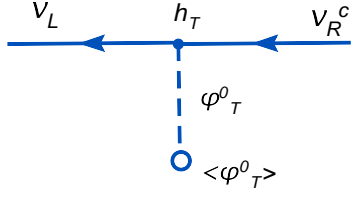}
\caption{Majorana mass term induced by the neutral component of a Higgs triplet field $\phi^0_T$ for an active neutrino.}
\label{mmterm}
\end{center}
\end{figure}

\begin{figure}[h]
\begin{center}
\includegraphics[height=2.5cm]{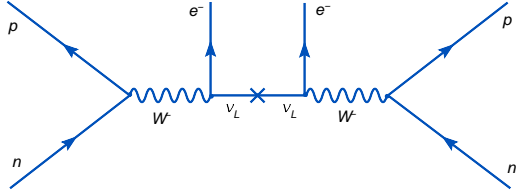}
\caption{Neutrinoless double beta decay}
\label{ndbd1}
\end{center}
\end{figure}

A Majorana mass term describes a transition between a  left-handed neutrino
and its CPT-conjugate right-handed antineutrino, as shown in Figure~\ref{majmass}. It can be viewed as the annihilation or creation of two neutrinos, and therefore violates lepton number by two units, $\Delta L=2$.
A Majorana mass term has the form
\begin{equation}
{-\cal L}=\frac{1}{2}m_{T}(\bar{\nu}_{L}\nu_{R}^{c}+\bar{\nu}_{R}^{c}\nu_{L})= 
\frac{m_{T}}{2}(\bar{\nu}_{L}C\overline{\nu_{L}}^{T}+h.c)=\frac{1}{2}m_{T}\bar{\nu}\nu,
\end{equation}
where $\nu=\nu_{L}+\nu_{R}^{c}$ is a self-conjugate two-component  (Majorana) field satisfying $\nu=\nu^{c}=C\bar{\nu}^{T}$. 
A Majorana mass for an active neutrino violates weak isospin by one unit, $\Delta I_{W}=1$, and can be generated either by the VEV of a Higgs triplet  (see Figure~\ref{mmterm})
or by a higher-dimensional operator involving two Higgs doublets (which could be generated, for
example, in a seesaw model).
A Majorana  $\nu$ is its own antiparticle and can mediate  neutrinoless double beta decay ($\beta \beta_{0\nu}$)~\cite{double1,double2},
in which two neutrons turn into two protons and two electrons, violating lepton number by two units, as shown in Figure~\ref{ndbd1}.

A sterile neutrino can also have a Majorana mass term of the form,
\begin{equation}
{-\cal L}=\frac{m_{S}}{2}[\bar{N}_{L}^{c}N_{R}+\bar{N}_{R}N_{L}^{c}].
\end{equation}
In this case, weak isospin is conserved, $\Delta I_{W}=0$, and thus $m_S$ can be generated by the VEV of a Higgs singlet\footnote{$m_S$ could also be generated in principle by a bare mass, but this is usually forbidden by additional symmetries in extensions of the SM.}.

\subsection{Mixed models}
If active and sterile neutrinos are both present, one can have both Dirac and Majorana mass terms simultaneously. For one
family, the Lagrangian has the form
\begin{equation}
{-\cal L}=\frac{1}{2}
\left(\begin{array}{cc}
\bar{\nu}_{L}^{0} & \bar{N}_{L}^{0c}\end{array}\right)
\left(\begin{array}{cc}
m_{T} & m_{D}\\
m_{D} & m_{S}
\end{array}\right)
\left(\begin{array}{c}\nu_{R}^{0c} \\ N_{R}^{0}\end{array}\right) + h.c., \label{mixed}
\end{equation}
where $0$ refers to weak eigenstates 
and the masses are
\[
\begin{array}{lllccl}
m_{T}:& |\Delta L|=2, && \Delta I_{W}=1&& (\mbox{Majorana}),\\
m_{D}:& |\Delta L|=0, && \Delta I_{W}=\frac{1}{2}&& (\mbox{Dirac}),\\
m_{S}:& |\Delta L|=2, && \Delta I_{W}=0&& (\mbox{Majorana}).
\end{array}
\]
Diagonalizing the matrix in~(\ref{mixed}) yields
two Majorana mass eigenvalues and two Majorana mass eigenstates, $\nu_{i}=\nu_{iL}+\nu_{iR}^c=\nu_{i}^{c}$, with 
$i=1,2$. The weak and mass bases are related by the unitary transformations
\begin{equation}
\begin{array}{lcccr}
\left(\begin{array}{c}\nu_{1L} \\ \nu_{2L} \end{array}\right)=
U_{L}^{\nu \dag} \left(\begin{array}{c}\nu_{L}^{0} \\ N_{L}^{0c}\end{array}\right), & & 
\left(\begin{array}{c}\nu_{1R}^c \\ \nu_{2R}^c \end{array}\right)=
U_{R}^{\nu \dag} \left(\begin{array}{c}\nu_{R}^{0c} \\ N_{R}^{0}\end{array}\right)
\end{array}.
\end{equation}
$U_L$ and $U_R$ are generally different for Dirac mass matrices, which need not be Hermitian. However, the general $2\times 2$ neutrino
mass matrix in~(\ref{mixed}) is symmetric because of~(\ref{adjoint}).
In our phase convention, in which $\nu_{iR}^c=C \overline{\nu_{iL}}^T$, this implies $U_L^\nu=U_R^{\nu \ast}$.

\subsection{Special Cases}
\begin{description}
\item[(a) Majorana:] $m_{D}=0$ in~(\ref{mixed}) corresponds to the pure Majorana case: the mass matrix is diagonal, with $m_{1}=m_{T}$, $m_{2}=m_{S}$, and
\[
\begin{array}{lcr}
\nu_{1L}=\nu_{L}^{0}, & & \nu_{1R}^{c}=\nu_{R}^{0c}, \\
\nu_{2L}=N_{L}^{0c}, & & \nu_{2R}^{c}=N_{R}^{0}.
\end{array}
\]
\item[(b) Dirac:]  $m_{T}=m_{S}=0$ is  the Dirac limit. There are formally two Majorana mass eigenstates, with
eigenvalues
$m_{1}=m_{D}$ and $m_{2}=-m_{D}$ and eigenstates 
\[
\begin{array}{lcr}
\nu_{1L}=\frac{1}{\sqrt{2}}( \nu_{L}^{0} +N_{L}^{0c}), & &  \nu_{2L}=\frac{1}{\sqrt{2}}( \nu_{L}^{0}-N_{L}^{0c}),\\
\nu_{1R}^{c}=\frac{1}{\sqrt{2}}( \nu_{R}^{0c}+N_{R}^{0}), & & \nu_{2R}^{c}=\frac{1}{\sqrt{2}}(\nu_{R}^{0c}-N_{R}^{0}).
\end{array}
\]
Note that $\nu_{1,2}$ are degenerate in the sense that $|m_1|=|m_2|$.  
To recover the Dirac limit, expand the mass term
\begin{equation}
\begin{array}{lcl}
{-\cal L}&=&\frac{m_{D}}{2}(\bar{\nu}_{1L}\nu_{1R}^{c}-\bar{\nu}_{2L}\nu_{2R}^{c})+h.c.\\
&=&\frac{m_{D}}{2}(\bar{\nu}_{L}^0N_{R}^0+\bar{N}_{R}^0\nu_{L}^0),
\end{array}
\end{equation}
which clearly conserves lepton number (i.e., there is no $\nu_L^0-N_L^{0c}$ or $\nu_R^{0c}-N_R^0$ mixing).
Thus, a Dirac neutrino can be thought of as two Majorana neutrinos, with maximal ($45^{\circ}$) mixing and
with equal and opposite masses. This interpretation is  useful in considering the Dirac limit of
general models.
\item[(c) Seesaw:]  $m_{T}=0$ and $m_{S} \gg m_{D}$ (e.g.,  
$m_{D}=O(m_{u},m_{e},m_{d})$ and $ m_{S}=O(M_{X})$, where $M_X\sim10^{14}$~GeV) is the seesaw limit~\cite{GRS,Yanagida,schechter},
with eigenstates and eigenvalues 
\[
\begin{array}{lcl}
\nu_{1L}\sim \nu_{L}^{0}-\frac{m_{D}}{m_{S}}N_{L}^{0c} \sim \nu_{L}^{0}, & -m_{1}\sim\frac{m^{2}_{D}}{m_{S}} \ll m_{D}, \\
\nu_{2L}\sim \frac{m_{D}}{m_{S}}\nu_{L}^{0}+N_{L}^{0c} \sim N_{L}^{0c},  & m_{2}\sim m_{S}.
\end{array}
\]
The seesaw mechanism is illustrated in Figure~\ref{seesaw12}.
\begin{figure}[h]
\begin{center}
\includegraphics[height=3cm]{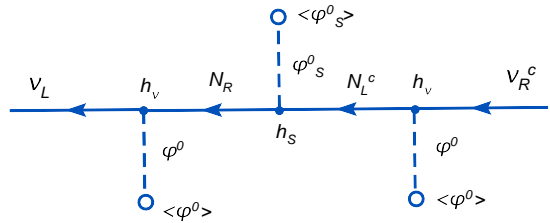}
\caption{The seesaw mechanism, in which a light active neutrino mixes with a very heavy Majorana neutrino.}
\label{seesaw12}
\end{center}
\end{figure}
\item[(d) Pseudo-Dirac:] this is a  perturbation on the Dirac case, with $m_{T}, \ m_{S}\ll m_{D}$. There is a small lepton number violation, and a small splitting between the magnitudes of the mass eigenvalues.
For example, $m_T=\epsilon,\ m_S=0$ leads to $|m_{1,2}|=m_D\pm\epsilon/2$.
\item[(e) Mixing:] The general case in which $m_{D}$ and $m_{S}$ (and/or $m_{T}$) are both small and comparable 
leads to non-degenerate Majorana mass eigenvalues and significant ordinary -- sterile ($\nu_{L}^{0}- N_{L}^{0c}$) mixing (as would be suggested if the  LSND results are confirmed). Only this and the
pseudo-Dirac cases allow such mixings.
\end{description}

As we have seen, the fermion mass eigenvalues can be negative or even complex. However, only the magnitude 
 is relevant for most purposes, and in fact fermion masses can always be made real and positive
by chiral transformations (e.g., redefining the phases of the $\nu_{iR}^c$). However, the relative signs (or phases)
may reappear in the $\beta \beta_{0\nu}$ amplitude or new (beyond the SM) interactions.

\subsection{Extension to Three Families}

It is straightforward to generalize~(\ref{mixed}) to three or more families, or even to the case of different numbers
of active and sterile neutrinos. For $F=3$ families, the Lagrangian is
\begin{equation}
{-\cal L}=\frac{1}{2}
\left(\begin{array}{cc}
\bar{\nu}_{L}^{0} & \bar{N}_{L}^{0c}\end{array}\right)
\left(\begin{array}{cc}
m_{T} & m_{D}\\
m_{D}^{T} & m_{S}
\end{array}\right)
\left(\begin{array}{c}\nu_{R}^{0c} \\ N_{R}^{0}\end{array}\right) + h.c.,
\label{matrixlag}
\end{equation}
where ${\nu}_{L}^{0}$ and  ${N}_{L}^{0c}$ are three-component vectors
\begin{equation}
\begin{array}{lcr}
\nu_{L}^{0}=\left(
\begin{array}{c}
\nu_{1L}^{0} \\
\nu_{2L}^{0} \\
\nu_{3L}^{0}
\end{array}
\right), & & 
N_{L}^{0c}=\left(
\begin{array}{c}
N_{1L}^{0c} \\
N_{2L}^{0c} \\
N_{3L}^{0c}
\end{array}
\right),
\end{array}
\end{equation}
and similarly for $\nu_{R}^{0c}$ and $ N_{R}^{0}$.  $m_{S}$, $m_{D}$ and $m_{T}$ are $3\times 3$ matrices, where $m_{S}=m_{S}^{T}$ and 
$m_{T}=m_{T}^{T}$ because of~(\ref{adjoint}). There are six Majorana mass eigenvalues and eigenvectors. The transformation to go from 
the weak to the mass basis is given by
\begin{equation}
\nu_{L}=U_{L}^{\nu \dag} \left(\begin{array}{c}\nu_{L}^{0} \\ N_{L}^{0c}\end{array}\right),
\label{unitrans}
\end{equation}
where $U_{L}^{\nu}$ is a $6 \times 6$ unitary matrix and $\nu_L$ is a six-component vector. The analogous transformation for the $R$ fields
involves $U_R=U_L^\ast$ because the $6 \times 6$ Majorana mass matrix is necessarily symmetric.

First consider the case in which there are only Dirac masses, which  is analogous to the quarks and charged leptons. 
The mass Lagrangian is 
\begin{equation}
{-\cal L}=\bar{\nu}_{L}^{0}m_{D} N_{R}^{0}+h.c.,
\end{equation}
where $m_D$ need not be Hermitian.
We may
change from the weak to the mass basis with a unitary transformation, 
\begin{equation}
{-\cal L}=\bar{\nu}_{L}\hat{m}_{D} N_{R},
\end{equation}
where $\hat{m}_{D}$ is the diagonal matrix of eigenvalues, $\hat{m}_{D}=U_{L}^{\nu \dag}m_{D}U_{R}^{\nu }={\rm diag}(m_{1},\ldots m_{3})$, and the eigenvectors are
\begin{equation}
\begin{array}{lcr}
\nu_{L}=U_{L}^{\nu \dag}\nu_{L}^{0}, && N_{R}=U_{R}^{\nu \dag} N_{R}^{0}
\end{array}.
\label{weakmass}
\end{equation}
If $m_{D}$ is not Hermitian, $U_{L,R}^\nu$ can be found by diagonalizing the Hermitian matrices $m_{D}m_{D}^{\dag}$ and $m_{D}^{\dag}m_{D}$, 
\begin{equation}
U_{L}^{\nu \dag}m_{D}m_{D}^{\dag}  U_{L}^{\nu }=U_{R}^{\nu \dag}m_{D}^{\dag}m_{D}  U_{R}^{\nu }=|\hat{m}_{D}|^{2}={\rm diag}(|m_{1}|^{2} \cdots |m_{3}|^{2}). \label{transf}
\end{equation}
The $U_{L,R}^{\nu}$ are not unique. If a given $U_{L}^{\nu}$ satisfies~(\ref{transf}), 
then so does \begin{equation}
\hat{U_{L}^{\nu}}=U_{L}^{\nu}K_{L}^\nu,
\end{equation}
where $K_{L}^\nu$ is a diagonal phase matrix\footnote{There is more freedom if there are degenerate eigenvalues.},
$
K_{L}^\nu={\rm diag}\left(\begin{array}{ccc}
e^{i\alpha_{1}} &
 e^{i\alpha_{2}}&
 e^{i\alpha_{3}}
\end{array}\right)$. This of course corresponds to the freedom to redefine the phases of the mass eigenstate fields.
Similar statements apply to $U_R^\nu$ and to the unitary transformations for the quark and charged lepton fields.

The weak charged current is given by 
\begin{equation}
J^{\mu}_{W} =2 \bar{e}^{0}_{L}\gamma^{\mu}\nu^{0}_{L}=2\bar{e}_{L}\gamma^{\mu}U_{L}^{e \dag} U_{L}^{\nu }\nu_{L},
\end{equation}
and $U_{L}^{e \dag} U_{L}^{\nu }$ is the lepton mixing matrix $V_{PMNS}$\footnote{This is for  Maki, Nakagawa, and Sakata~\cite{mns} in 1962 and Pontecorvo~\cite{pontecorvo} in 1968.}. A general unitary $3\times 3$ matrix involves 9 parameters: 3 mixing angles
and 6 phases. However, five of the phases in $V_{PMNS}$ (and similarly in $V_{CKM}$) are unobservable in the Dirac case
because they depend on the relative phases of the left-handed neutrino and charged lepton fields, i.e., they
can be removed by an appropriate choice of $K_L^\nu$ and the analogous $K_L^e$,
so there is only one physical CP-violating phase. The analogous 
$K_R^{\nu, e}$ in $U_R^{\nu,e}$ can then be chosen to make the mass eigenvalues real and positive. The 
$K_R$ phases
are unobservable unless there are new BSM interactions involving the $R$ fields.

The diagonalization of a Majorana mass matrix is similar, except for the constraint $U_L^\nu=U_R^{\nu \ast}$.
This implies that $K_L^\nu$ and $K_R^\nu$ cannot be chosen independently (because the $L$ and $R$ mass
eigenstates are adjoints of each other). One must then use the freedom in $K_L^\nu$ to make the mass eigenvalues
real and positive, so that one cannot remove as many phases from the leptonic mixing matrix. For three light
mass eigenstates
(e.g., for three active neutrinos with $m_T\ne 0, m_D=0$; or  for
the seesaw limit of the $6\times 6$ case), this implies that there are two additional ``Majorana'' phases,
i.e., $V_{PMNS}= \hat{V}_{PMNS} \times {\rm diag}\left(\begin{array}{ccc}
e^{i\beta_{1}} &
 e^{i\beta_{2}}&
 1
\end{array}\right)$, where $ \hat{V}_{PMNS}$ has the canonical form with one phase, and $\beta_{1,2}$ are
CP-violating relative phases associated with the Majorana neutrinos. $\beta_{1,2}$ do not affect neutrino
oscillations, but can in principle affect $\beta \beta_{0\nu}$ and new interactions.

Let us conclude this section with a few comments.
\begin{itemize}
\item The LSND oscillation results would, if confirmed, strongly suggest the existence of
very light sterile neutrinos which mix with active neutrinos of the same chirality~\cite{bbns2}. The
pure Majorana and pure Dirac cases do not allow any such mixing (sterile neutrinos
are not even required  in the Majorana case). The seesaw limit only has very heavy sterile
neutrinos and negligible mixing. Only the general and pseudo-Dirac limits allow significant
ordinary-sterile mixing, but in these cases one must find an explanation for two very small
types of masses.
\item There is no distinction between Dirac and Majorana neutrinos except by their masses
(or by new interactions). As the masses go to zero, the active components reduce to standard
active Weyl spinors in both case. There are additional sterile Weyl spinors in the massless
limit of the Dirac case, but these decouple from the other particles.
\item One can ignore $V_{PMNS}$ in processes for which the neutrino masses are too small
to be relevant. In that limit, the neutrino masses are effectively degenerate (with vanishing mass)
and one can simply work in the weak basis.
\end{itemize}

\subsection{Models of Neutrino Mass}
There are an enormous number of models of neutrino mass~\cite{massmodels}.

Models to generate Majorana masses are most popular, because no Standard Model
gauge symmetry forbids them. (However, models descended from underlying string
constructions may be more restrictive because of the underlying symmetries and selection
rules~\cite{strings}.)
Models constructed to yield small Majorana masses include:
the ordinary (type I) seesaw~\cite{GRS,Yanagida,schechter},
often combined with additional family and grand unification symmetries;
models with heavy Higgs triplets (type II seesaw)~\cite{schechter,typeii,typeii2};
TeV (extended) seesaws~\cite{extended}, with $m_\nu \sim m^{p+1}/M^p, \ \ p>1$, e.g., with $M$ in the TeV range;
radiative masses (i.e., generated by loops)~\cite{zee};
supersymmetry with $R$-parity violation, which may include loop effects;
supersymmetry with mass generation by terms in the K\" ahler potential;
anarchy (random entries in the mass matrices);
and large extra dimensions (LED),  possibly combined with one of the above. 

Small Dirac masses may be due to:
 higher dimensional
 operators (HDO) in intermediate scale models (e.g., associated with an extended 
 $U(1)'$ gauge symmetry~\cite{uprm} or supersymmetry breaking); large intersection areas in intersecting brane models 
 or large extra dimensions, from volume suppression if $N_R$ propagates in the bulk~\cite{largedim}.

Simultaneous small Dirac and Majorana masses, as motivated by LSND,  may be due, e.g.,  to 
HDO~\cite{sterile2} or to sterile neutrinos
from a ``mirror world''~\cite{sterile1}.

There are also many ``texture'' models~\cite{altarelli}, involving specific guesses about the form of the 
$3 \times 3$ neutrino mass matrix
or the Dirac and Majorana matrices entering seesaw models. These are often studied 
in connection with models also involving quark
and charged lepton mass matrices, 
such as grand unification (GUTs), family symmetries, or left-right symmetry.

\section{Laboratory and Astrophysical Constraints on Neutrino Counting and Mass}
\subsection{Laboratory Limits}
The most precise measurement of the number of light ($m_\nu < M_Z/2$) active neutrino types 
and therefore the number of associated fermion families comes from the invisible $Z$ width $\Gamma_{inv}$, obtained by subtracting the observed width into quarks and charged leptons
from the total width from the lineshape. The number of  effective  neutrinos $N_{\nu}$ is given by
\begin{equation}
N_{\nu}=\frac{\Gamma_{inv}}{\Gamma_{l}}\left(\frac{\Gamma_{l}}{\Gamma_{\nu}}\right)_{SM},
\label{numbernu}
\end{equation}
where $(\Gamma_{l}/\Gamma_{\nu})_{SM}$ is the SM expression for the
ratio of widths into a charged lepton and a single active neutrino, introduced to reduce the model dependence. The experimental value  is $N_{\nu}=2.984 \pm 0.008$~\cite{ewreview}, excluding the
possibility of a fourth family unless the neutrino is very heavy. Other unobserved particles
from $Z$ decay would also give a positive contribution to  $N_{\nu}$. For example,
the decay $Z\longrightarrow M S$ in models with spontaneous lepton number violation and a Higgs triplet~\cite{GRmodel},
where $M$ is a Goldstone boson (Majoron) and $S$ is a light scalar, would yield a contribution of
2 to $N_{\nu}$ and is therefore excluded.

Kinematic laboratory measurements  on  neutrino masses are relatively weak~\cite{ywang}. For the $\tau$-neutrino the most stringent limit is $m_{\nu_{\tau}}<18.2$~MeV, which comes from the decay channel $\tau\rightarrow 5 \pi + \nu_{\tau}$. For the $\mu$-neutrino the most sensitive measurement gives $m_{\nu_{\mu}}<0.19$~MeV from  $\pi\longrightarrow\mu\nu_\mu$ decay\footnote{The limits
on $m_{\nu_\tau}$ and $m_{\nu_\mu}$ are now obsolete because of oscillation and cosmological constraints.}. The analysis of tritium beta decay at low energies gives a limit on the mass of  the $e$-neutrino of $m_{\nu_{e}}\sim 2.8$~eV.
Including mixing, this should be interpreted as a limit on
$m_\beta \equiv \sqrt{\Sigma_i |V_{ei}|^2 |m_i|^2}$, where $V=V_{PMNS}$. The latter should be improved
in the future to a sensitivity of around 0.2~eV by the KATRIN experiment. 

For Majorana masses, the amplitude for neutrinoless double beta decay~\cite{double1,double2} is $A\sim A_{nuc}m_{\beta\beta}$, where 
$A_{nuc}$ contains the nuclear matrix element and $m_{\beta\beta}\equiv  \sum_{i} (V_{ei})^{2} m_{i} $
is the effective Majorana mass in the presence of mixing between light Majorana neutrinos\footnote{There can also be other BSM contributions to the decay, such as heavy Majorana neutrinos or $R$-parity violating effects
in supersymmetry.}. $m_{\beta\beta}$ is just the $(1,1)$ element of $m_T$ or of the effective
Majorana mass matrix in a seesaw model. It involves  the square of  $V_{ei}$ rather than the
absolute square, and is therefore sensitive in principle to the Majorana phases (though this is difficult in practice). The expression allows for the possibility of cancellations between
different mass eigenstates\footnote{In the convention for the field phases  that the masses are real and positive the cancellations are due to the phases in $V_{ei}$. Alternatively, one can use the convention that the mass eigenvalues may be
negative or complex.}, and in fact shows why $\beta \beta_{0\nu}$ vanishes for a Dirac neutrino, which
can be viewed as two canceling Majorana neutrinos. At present,
$|m_{\beta\beta}|<(0.35-1)$~eV, where the range is from the theoretical uncertainty in $A_{nuc}$. Members of one experiment claimed an  observation of $\beta \beta_{0\nu}$, which corresponds to $m_{\beta \beta}\sim 0.39$~eV, but this has not been confirmed.
Future  experiments should be sensitive to $m_{\beta \beta}\sim (0.1-0.2\mbox{ eV})$.

\subsection{Cosmological Constraints}

The light nuclides $^{4}He$, $D$, $^{3}He$, $^{7}Li$ were synthesized in the first thousand seconds in the early evolution the 
universe~\cite{Dolgov:2002wy,Fields:2004zw}, corresponding to temperatures from 1~MeV to 50~keV. At temperatures above the freezeout 
temperature $T_f \sim {\rm few}$~MeV, the neutron to proton ratio was kept in equilibrium by the reactions
\begin{equation}
\begin{array}{ccc}
n + \nu_{e} & \leftrightarrow & p + e^{-}, \\
\\
n + e^{+} & \leftrightarrow & p + \bar{\nu}_{e}.
\end{array}
\label{reactions}
\end{equation}
For $T< T_f$ their rates became slow compared to the expansion rate of the universe, and the neutron to proton ratio, $n/p$, 
froze at a constant (except for neutron decay) value $\frac{n}{p}=\exp(-\frac{m_{n}-m_{p}}{T_{f}})$. Most of the neutrons were incorporated
into $^{4}He$, so that the primordial abundance (relative to hydrogen) can be predicted\footnote{There is also a weak dependence on 
the baryon density relative to photons, which is determined independently by the $D$ abundance and by the cosmic microwave background (CMB)
anisotropies.} in terms of $T_f$. $T_f$ is predicted by comparing the reaction rate for the processes in Eq.~(\ref{reactions}), 
$\Gamma \sim G_{F}^{2}T^{5}$, and the Hubble expansion rate, $H=1.66 \sqrt{g_{\star}}\frac{T^{2}}{M_{\rm Pl}} \sim \sqrt{g_{\star}}T^{2}$,
where $M_{\rm Pl}$ is the Planck scale.
Therefore,  $T_{f}\sim g_{\star}^{1/6}$, where $g_{\star}$ counts the number of relativistic particle species, determining 
the energy density in radiation. It is given by $g_{\star}=g_{B}+\frac{7}{8}g_{F}$, where $g_{F}=10+2\Delta N_{\nu}'$, and where 
the $10$ is due to $3 \nu+3\bar{\nu}+2 \mbox{ helicities each of }e^{\pm}$. $\Delta N_{\nu}'$ is the effective number of additional 
neutrinos present at $T\gsim T_f$. It includes new active neutrinos with masses\footnote{There would be an enhanced contribution from 
a $\nu_\tau$ in the 1-20~MeV range, which was once important in constraining its mass, or a reduced contribution from $\nu_\tau$ decay.} 
$\lsim 1$~MeV, and also light sterile neutrinos, which could be produced by mixing with active neutrinos for a wide range of mixing 
angles\footnote{The effects of light sterile neutrinos can be avoided in variant scenarios or compensated by a large
$\nu_e-\bar{\nu}_e$ asymmetry~\cite{ichep}.}. It does not include the right-handed sterile components of light Dirac neutrinos, 
which would not have been produced in significant numbers unless they couple to BSM interactions. The prediction of the primordial mass 
abundance of $^{4}He$ relative to $H$ is $\sim 24$\% for $\Delta N_\nu'=0$. There is considerable uncertainty in the observational value, 
but most estimates yield $\Delta N_{\nu}'< 0.1-1$.

Neutrinos with masses in the eV range would contribute hot dark matter to the universe~\cite{Dolgov:2002wy}. They would close the universe,
$\Omega_{\nu}=1$, for the sum of masses  of the light neutrinos (including sterile neutrinos, weighted by their abundance relative to
active neutrinos) $\Sigma\equiv \Sigma_{i} |m_{i}|\sim 35$~eV. However, even though some 30\% of the energy density is believed to be in 
the form dark matter, most of it should be cold (such as weakly interacting massive particles) rather than hot (neutrinos), because 
the latter cannot explain the formation of smaller scale structures during the lifetime of the universe. The Wilkinson Microwave Anisotropy
Probe (WMAP), together with the Sloan Digital Sky Survey (SDSS), the Lyman alpha forest (Ly$\alpha$), and other observations, leads 
to strong constraints on $\Sigma \lsim 1$~eV~\cite{Spergel:2003cb}, with the most stringent claimed limit of 0.42~eV~\cite{cosmological}. 
Using future Planck data, it may be possible to extend the sensitivity down to $0.05-0.1$~eV, close to the minimum value 
$0.05\mbox{ eV} \sim \sqrt{|\Delta m^2_{\rm atm}|}$ allowed by the neutrino oscillation data. 
However, there are significant theoretical uncertainties.

\subsection{Neutrino Oscillations}
Neutrino oscillations can occur due to the mismatch between weak and mass eigenstates, 
and are analogous to the time evolution of a quantum system which is not in an energy eigenstate, or
a classical coupled oscillator in which one starts with an excitation that is not a normal
mode.

Consider a system in which there are only  two neutrino flavors, e.g., $\nu_e$ and $\nu_\mu$. Then
\begin{equation}
\begin{array}{l}
|\nu_{e}\rangle =|\nu_{1}\rangle \cos \theta + |\nu_{2}\rangle \sin \theta, \\
|\nu_{\mu}\rangle =-|\nu_{1}\rangle \sin \theta + |\nu_{2}\rangle \cos \theta,
\end{array}
\end{equation}
where $\theta$ is the neutrino mixing angle. Suppose that at initial time, $t=0$, we create a pure weak
eigenstate (as is typically the case), such as $\nu_\mu$ from the decay $\pi^+\longrightarrow
\mu^+\nu_\mu$, i.e.,
 $|\nu(0)\rangle =|\nu_{\mu}\rangle $. Since $|\nu_{\mu}\rangle $ is a superposition of mass
 eigenstates, each of which propagates with its own time dependence, then at a later time
 the state may have oscillated into $|\nu_{e}\rangle$, which can be identified by its interaction,
 e.g., $\nu_e n \longrightarrow e^-p$. To quantify this, after a time $t$ the state $|\nu(0)\rangle$
 will have evolved into
\begin{equation}
|\nu(t)\rangle =-|\nu_{1}\rangle \sin \theta e^{-iE_{1}t}+ |\nu_{2}\rangle \cos \theta e^{-iE_{2}t},
\end{equation}
where $E_{1}=\sqrt{p^{2}+m_{1}^{2}} \sim p+\frac{m_{1}^{2}}{2p}$ and $E_{2}=\sqrt{p^{2}+m_{2}^{2}} \sim p+\frac{m_{2}^{2}}{2p}$, and we have assumed that the neutrino is highly relativistic, $p \gg m_{1.2}$,
as is typically the case. After traveling a distance $L$ the oscillation probability, i.e., the 
probability for the neutrino to interact as a $\nu_e$, is
\begin{equation}
\begin{array}{lcl}
P_{\nu_{\mu}\rightarrow \nu_{e}}{(L)}&=&|\langle \nu_{e}|\nu(t)\rangle |^{2} \\
&=&\sin^{2} \theta \cos^{2} \theta \ |-e^{-iE_{1}t}+e^{-iE_{2}t}|^{2}\\
&=&\sin^{2} 2\theta \sin^{2}  \left( \frac{\Delta m^{2}L}{4E}\right)=\sin^{2} 2\theta \sin^{2}  
\left( \frac{1.27 \Delta m^{2}({\rm eV}^{2})L({\rm km})}{E({\rm GeV})}\right),
\end{array}
\end{equation}
 where $L \sim t$, $E=p$ and $\Delta m^2=m_{2}^{2}-m_{1}^{2}$. The probability for 
 the state to remain a $\nu_{\mu}$ is $P_{\nu_{\mu}\rightarrow \nu_{\mu}}{(L)}=1-P_{\nu_{\mu}\rightarrow \nu_{e}}{(L)}$. 
 
 There are two types of oscillation searches:
\begin{enumerate}
\item Appearance experiments, in which one  looks for the appearance of a new flavor, e.g., of $\nu_{e}$ or $\nu_\tau$ (i.e., of $e^{-}$ or $\tau^-$) in an initially pure $\nu_{\mu}$ beam.
\item Disappearance experiments, in which one looks for a change in, e.g., the  $\nu_{\mu}$ flux as a function of $L$ and $E$.
\end{enumerate}

Even with more than two types of neutrino, it is often a good approximation to use the two neutrino
formalism in the analysis of a given experiment. However, a more precise or general analysis
should take all three neutrinos into account\footnote{If there are light sterile neutrinos which mix with
the active neutrinos one should generalize to include their effects as well.}.
The general lepton mixing for three families contains three angles $\theta_{12},\ \theta_{13},$
and $\theta_{23}$; one Dirac CP-violating phase $\delta$;  and two Majorana phases
$\beta_{1,2}$. (The latter do not enter the oscillation formulae.) In the basis in which
the charged leptons are mass eigenstates, the neutrino mass eigenstates $\nu_{i}$ and the weak eigenstates $\nu_{a}=(\nu_{e},\nu_{\mu},\nu_{\tau})$ are related by a unitary transformation $\nu_{a}=V_{ai}\nu_{i}$, where $V$is  the $3 \times 3$ lepton mixing matrix $V_{PMNS}$, which is parametrized as 
\begin{equation}
\begin{array}{lcl}
V_{PMNS}&=&\left[
\begin{array}{ccc}
c_{12}c_{13} & s_{12}c_{13} & s_{13}e^{-i\delta} \\
-s_{12}c_{23}-c_{12}s_{23}s_{13}e^{i\delta} & c_{12}c_{23}-s_{12}s_{23}s_{13}e^{i\delta} 
&  s_{23}c_{13} \\
s_{12}s_{23}-c_{12}c_{23}s_{13}e^{i\delta} & -c_{12}s_{23}-s_{12}c_{23}s_{13}e^{i\delta} &  c_{23}c_{13} 
\end{array}
\right]\\
\\
&& \times {\rm diag}(e^{i\beta_{1}/2},e^{i\beta_{2}/2},1),
\end{array}
\end{equation}
where $c_{ij}=\cos \theta_{ij}$ and $s_{ij}=\sin \theta_{ij}$.
The oscillation probability for  $\nu_a \longrightarrow \nu_b$ after a distance $L$ is then
\begin{equation}
P_{\nu_{a}\rightarrow \nu_b}{(L)} \underbrace{=}_{a\ne b} \Sigma_j \left| V_{aj}V_{bj}\right|^2 
 + {\rm Re}\ \Sigma_{i\ne j}V_{ai}V_{bi}^\ast V_{aj}^\ast V_{bj} e^{-i \Delta_{ij} L/2E},
\end{equation}
where $\Delta_{ij}= m_i^2-m_j^2$.

\section{Atmospheric Neutrinos}
\begin{figure}[h]
\begin{center}
\includegraphics[height=7cm]{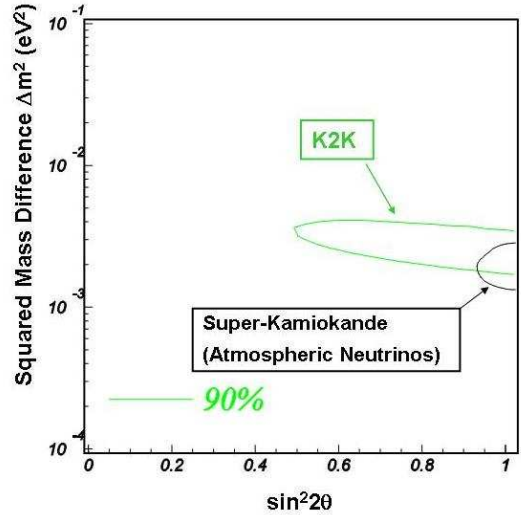}
\caption{Contours enclosing the 90\% confidence regions for two-flavor ($\nu_\mu\leftrightarrow \nu_\tau$) neutrino oscillation parameters for the K2K results, compared to results from Super-Kamiokande atmospheric neutrinos (from~\cite{superkamioka}).}
\label{kamioka}
\end{center}
\end{figure}

Many experiments have searched for neutrino oscillations at reactors, accelerators, and from astrophysical sources~\cite{nuosc}. 
Although the first indications of an effect involved the Solar neutrinos, the first unambiguous evidence came from the oscillations of 
atmospheric neutrinos. Atmospheric neutrinos are the product of pion and muon decays, which are produced in the upper layers of 
the atmosphere due to the interaction of  primary cosmic rays. The data from the Kamiokande and Super-Kamiokande water 
$\check{\rm C}$herenkov detectors 
indicated the disappearance of $\mu$-neutrinos. This was first seen in the ratio of the $\nu_\mu/\nu_e$ fluxes, and later confirmed by the
zenith angle distribution of $\nu_\mu$ events. Results from other experiments such as MACRO and
Soudan confirm the results, as does the recent long-baseline K2K experiment involving neutrinos
produced at the KEK lab and observed in the Super-Kamiokande detector
(Figure~\ref{kamioka}). The details of the observations have now established that
the dominant effect is indeed oscillations of the  $\nu_{\mu}$ into\footnote{Oscillations into
$\nu_{e}$ or a sterile neutrino as the dominant  effect are excluded  by the Super-Kamiokande data and reactor experiments.} $\nu_{\tau}$, with near-maximal mixing ($\sin^{2}2\theta_{23} > 0.92$), and the difference  between the mass squares of the two mass eigenstates of order $|\Delta m^{2}_{atm}| \sim 2\times 10^{-3}$~eV. 

\section{Solar Neutrinos}
\begin{figure}[h]
\begin{center}
\includegraphics[height=8cm]{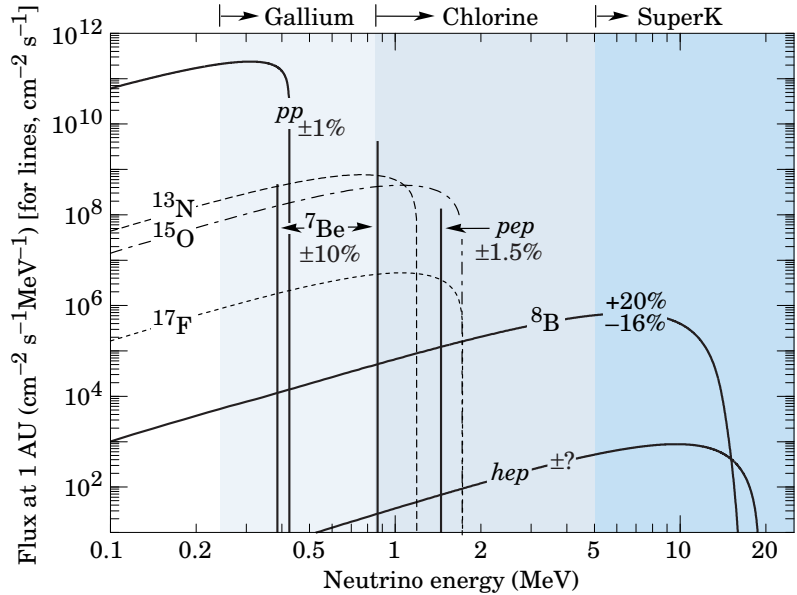}
\caption{Total predicted flux vs.\ energy of Solar neutrinos in the Standard Solar Model. 
The ranges of sensitivity of the various experiments is also shown, the SNO range being similar to SuperK (from~\cite{jnbahcallweb}).}
\label{tfn}
\end{center}
\end{figure}

The first indications~\cite{solarrev} of neutrino oscillations were from the low number of $\nu_e$ events observed
in the Davis Solar neutrino chlorine experiment, compared with the predictions
of the  Standard Solar Model (SSM)~\cite{Bahcall:2004pz}, shown in  Figure~\ref{tfn}. 
The deficit was later confirmed (and that the observed events do indeed come from the Sun) in the Kamiokande and Super-Kamiokande water
$\check{\rm C}$herenkov experiments. These observed the high energy $^8B$ neutrinos by the reaction
$\nu e^- \longrightarrow \nu e^-$, for which the cross section for oscillated $\nu_{\mu, \tau}$ by the neutral current is about 1/6 
that for $\nu_e$. The gallium experiments GALLEX, SAGE, and (later) GNO were sensitive to the entire Solar spectrum and also showed 
a deficit. Comparing the depletions in the various types of experiments, which constituted a crude measurement of the spectrum, showed 
that neutrino oscillations of $\nu_e$ into $\nu_{\mu, \tau}$ or possibly a sterile neutrino were strongly favored over any plausible
astrophysical uncertainty in explaining the results~\cite{bhl}. Combining with information on the spectrum and time dependence of 
the signal (which could vary between day and night due to matter effects in the Earth) zeroed in on two favored regions for the oscillation
parameters, one with small mixing angles (SMA) similar to the quark mixing, and one with large mixing angles (LMA), and with 
$\Delta m^2 \sim 10^{-5}-10^{-4}\mbox{ eV}^2$. The LMA solution allowed oscillations to $\nu_{\mu,\tau}$ but not to sterile neutrinos as 
the primary process (they were differentiated by the sensitivity to $\nu_{\mu,\tau}$ in the water $\check{\rm C}$herenkov experiments), 
while the SMA solution allowed both\footnote{Big bang nucleosynthesis also disfavored the LMA solution for sterile neutrinos.}.

\begin{figure}[h]
\begin{center}
\includegraphics[height=7cm]{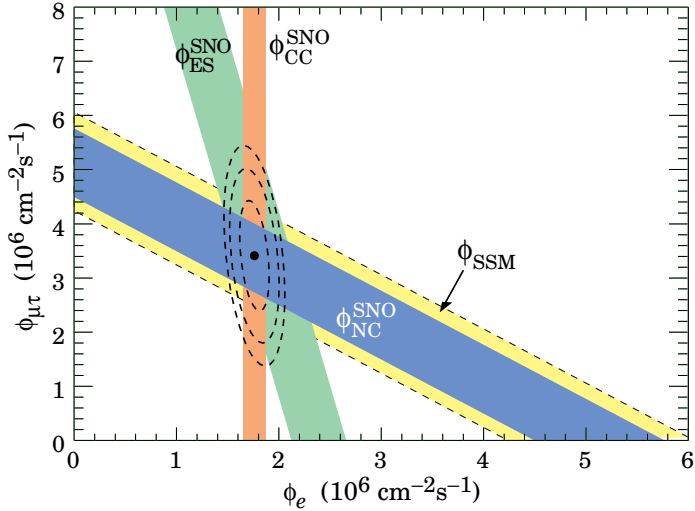}
\caption{Fluxes of $^{8}B$ Solar neutrinos deduced from the SNO charged current (CC), 
neutral current (NC) and electron scattering (ES) reactions, and the prediction of the
Standard Solar Model (from~\cite{snolab}).}
\label{snoflux}
\end{center}
\end{figure}

The situation was clarified by the Sudbury Neutrino Observatory (SNO) experiment~\cite{snolab}. SNO uses heavy water $D_{2}O$, 
and can measure three reactions,
\[
\begin{array}{c}
\nu_e + D \rightarrow e^{-} +p +p, \\
\nu_{e,\mu,\tau} + D \rightarrow \nu_{e,\mu,\tau} + p+ n,\\
\nu_{e,\mu,\tau} + e^{-}\rightarrow \nu_{e,\mu,\tau} + e^{-}. 
\end{array}
\]
In the first (charged current) reaction, the deuteron breakup can be initiated only by the $\nu_{e}$, while the second (neutral current) 
reaction can be initiated by neutrinos of all the active flavors with equal cross section. SNO could therefore determine the $\nu_e$ flux 
arriving at the detector and the total flux into active neutrinos (which would equal the initially produced $\nu_e$ flux if there
are no oscillations into sterile neutrinos) separately, as in Figure~\ref{snoflux}. SNO observed that  the total flux is around three times
that of $\nu_e$, establishing that oscillations indeed
take place, and that the total flux agrees well with the predictions of the SSM.
They also verified the LMA solution and that the Solar mixing angle $\theta_{12}$ is large
but not maximal, i.e., $\sin^{2}2\theta_{12}\sim 0.8$.

The Solar neutrino conversions are actually not due just to the vacuum oscillations as the
neutrinos propagate from the Sun to the Earth. One must also take into account the
coherent forward scattering~\cite{wolfenstein, mikheyev} of the neutrinos from matter in the Sun (and in the Earth at night),
which introduces the analog of an index of refraction. This is of order $G_F$ rather than
$G_F^2$, and it distinguishes $\nu_e$, $\nu_{\mu,\tau}$, $\bar{\nu}_e$, $\bar{\nu}_{\mu,\tau}$, and sterile neutrinos $\nu_s$ because of their different weak interactions.
The propagation equation in electrically neutral matter is
\[
i \frac{d}{dt}\left(
\begin{array}{c}
\nu_{e} \\
\nu_{\mu,\tau,s} 
\end{array}\right)=\left(\begin{array}{cc}
-\frac{\Delta m^{2}}{2E}\cos 2\theta + \sqrt{2}G_{F}n & \frac{\Delta m^{2}}{2E}\sin 2\theta\\
 \frac{\Delta m^{2}}{2E}\sin 2\theta\ & \frac{\Delta m^{2}}{2E}\cos 2\theta - \sqrt{2}G_{F}n 
\end{array}\right),
\]
where
\[
n=\left\{\begin{array}{lcr}
n_{e} &\mbox{ for}&  \nu_{e}\rightarrow \nu_{\mu,\tau}, \\
n_{e}-\frac{1}{2}n_{n} &\mbox{ for}&  \nu_{e} \rightarrow \nu_{s},
\end{array}\right.
\]
and where $n_e (n_n)$ is the number density of electrons (neutrons). This reduces to the
vacuum oscillation case (with parameters  $\theta$ and $\Delta m^2$) for $n=0$. 
Under the right conditions, the matter effect can greatly enhance the transitions.
In particular, if the Mikheyev-Smirnov-Wolfenstein (MSW)  resonance condition $\frac{\Delta m^{2}}{2E}\cos 2\theta=\sqrt{2}G_{F}n$  is satisfied, the diagonal elements vanish and even small vacuum mixing angles lead to
a maximal effective mixing angle. In practice $n$ decreases as the neutrinos propagate from the
core of the Sun where they are produced to the outside. For the LMA parameters, the higher 
energy Solar neutrinos all encounter a resonance layer. It is hoped that a future precise
experiment sensitive to the lower energy $pp$ neutrinos will observe the transition between
the vacuum and MSW dominated regimes.

Recently, the KamLAND long-baseline reactor experiment~\cite{kamland} in Japan has beautifully confirmed the LMA solution, free of any astrophysical uncertainties, by observing
a depletion and spectral distortion of $\bar{\nu}_e$ produced by power reactors located $O(100)$~km away.
The combination of KamLAND and the Solar experiments also limits the amount of oscillations
into sterile neutrinos that could accompany the dominant oscillations into $\nu_{\mu, \tau}$.

\section{Neutrino Oscillation Patterns}
The neutrino oscillation data is sensitive only to the mass-squared differences. 
The atmospheric and Solar neutrino mass-squares are given by
\[
\begin{array}{c}
|\Delta m^{2}_{32}|\equiv |\Delta m^{2}_{atm}| \sim 2 \times 10^{-3}\mbox{ eV}^{2}, \\
\Delta m^{2}_{21} \equiv \Delta m^{2}_{solar} \sim8 \times 10^{-5}\mbox{ eV}^{2}.
\end{array}
\]
Vacuum oscillations depend only on the magnitude of $\Delta m^2$, so the sign of $\Delta m^{2}_{atm}$
is not known, while MSW (matter) effects establish that  $\Delta m^{2}_{solar}>0$.
The respective mixing angles are $\sin^{2}2\theta_{23} > 0.92$ (90\%), consistent with maximal; and $\sin^{2}2\theta_{12}\sim 0.8$, i.e., $\tan^2 \theta_{12}, =0.40^{+0.09}_{-0.07}$, which is 
 large but not maximal. On the other hand, the third angle is small,
 $\sin^2 \theta_{13} < 0.03$ (90\%) from short-baseline ($\sim 1$~km) reactor disappearance limits,
 especially CHOOZ~\cite{ywang}.

The LSND experiment~\cite{LSND} at Los Alamos has claimed evidence for oscillations, especially $\bar{\nu}_\mu \longrightarrow \bar{\nu}_e$,
with $|\Delta m^2_{\rm  LSND}| \gsim 1\mbox{ eV}^2$ and small mixing. This has not been confirmed
by the KARMEN experiment, but there is a small parameter region allowed by both.
The  Fermilab MiniBooNE experiment is currently running, and should be able to confirm or
exclude the LSND results. If LSND is confirmed, the most likely explanation would be that there
is mixing of a fourth neutrino with the three known ones. This would have to be sterile because the
invisible $Z$ width excludes a fourth light active neutrino.

\subsection{Three Neutrino Patterns}
\begin{figure}[h]
\begin{center}
\includegraphics[height=4cm]{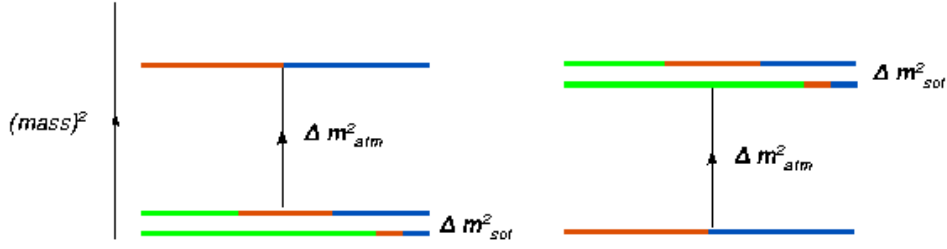}
\caption{Three neutrino mass-squared patterns. The $\nu_{e}$ fraction of each eigenstate is in green (light gray in black \& white 
printing), the $\nu_{\mu}$ fraction is indicated in red (dark gray), and the $\nu_{\tau}$ fraction in dark blue (black).
The left (right) figure is the normal (inverted) hierarchy.}
\label{3masspattern}
\end{center}
\end{figure}

If the LSND results are not confirmed, then the remaining data can be described by oscillations
amongst the three light active neutrinos.
Let us choose a phase convention in which the mass eigenvalues are real and positive,
and label the states such that $m_1<m_2$ is responsible for the Solar oscillations,
while the atmospheric oscillations are due to the $32$ mass-squared difference.
Since neither the sign of $\Delta m^{2}_{32}$ nor the absolute scale of the masses is known,
there are several possible patterns  for the masses. These include:
\begin{description}
\item[(1)] The normal (or ordinary) hierarchy, i.e., $m_{1}\ll m_{2}\ll m_{3}$, as in the left diagram in Figure~\ref{3masspattern}. In 
this case, $m_{3}\simeq \sqrt{\Delta m^{2}_{atm}}\sim 0.04-0.05$~eV, $m_{2}\simeq\sqrt{\Delta m^{2}_{sol}}\sim 0.009$~eV, and $m_1 \sim 0$.
\item [(2)] The inverted (or  quasi-degenerate) hierarchy, i.e., $m_{1}\simeq  m_{2}
\simeq \sqrt{\Delta m^{2}_{atm}}\sim 0.04-0.05\mbox{ eV} \gg m_3$, as in the right diagram in Figure~\ref{3masspattern}.
\item [(3)] The degenerate case, i.e., $m_{1}\simeq  m_{2}\simeq m_{3}$, with small
splittings responsible for the oscillations.
\end{description}
Of course, these are only limiting cases. One can interpolate smoothly from cases (1) and (2)
to the degenerate case by increasing the mass of the lightest neutrino.

\subsection{Four Neutrino Patterns}  
\begin{figure}[h]
\begin{center}
\includegraphics[height=4cm]{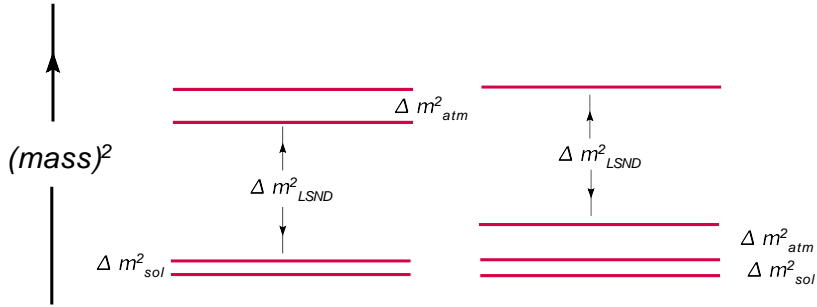}
\caption{The $2+2$ (left) and $3+1$ (right) four neutrino mass-squared patterns.
One can invert the sign of $\Delta m^{2}_{LSND}$ and (for the $3+1$ case) $\Delta m^{2}_{atm}$.}
\label{4masspattern}
\end{center}
\end{figure}

If the LSND data is confirmed, then there is most likely at least one light sterile neutrino. In the case of four neutrinos the  mass-squared difference for LSND is $\Delta m^{2}_{LSND}\sim 1$~eV$^2$.
In the  $2+2$ pattern in  Figure~\ref{4masspattern} the sterile component must be
mixed in significantly with either the Solar or atmospheric pair.
However, it is well established that neither the 
Solar nor the atmospheric oscillations are  predominantly into sterile states, and this scheme
is excluded. In the $3+1$ schemes a predominantly sterile state is separated from
the three predominantly active states. This is consistent with the Solar and
atmospheric data, but excluded when reactor and
accelerator disappearance limits are incorporated~\cite{fournu,Schwetz:2003pv}.
However, some 5 $\nu$ (i.e., $3+2$) patterns involving mass splittings around 1~eV$^2$ and
20~eV$^2$ are more successful~\cite{Sorel:2003hf}. All of these run into cosmological
difficulties, although there are some  (highly speculative/creative)
loopholes~\cite{ichep}. 

\section{Conclusions}
Non-zero neutrino mass is the first necessary extension of the Standard Model. Most extensions of the SM predict non-zero neutrino masses at some level, so it is difficult to determine their origin.
Many of the promising mechanisms involve very short distance scales, e.g., associated with grand unification or string theories.  There are many unanswered questions. These include:
\begin{itemize}
\item Are the neutrinos Dirac or Majorana? Majorana masses, especially if associated with
a seesaw or Higgs triplet, would allow the possibility
of leptogenesis, i.e., that the observed baryon asymmetry was initially generated as a lepton
asymmetry associated with neutrinos, and later converted to a baryon (and lepton)
asymmetry by nonperturbative electroweak effects. The observation of neutrinoless
double beta decay would establish Majorana masses (or at least $L$ violation), but
foreseeable experiments will only be sensitive to the inverted or degenerate spectra.
If the neutrinos are Dirac, this would suggest that additional TeV scale symmetries or string symmetries/selection rules are forbidding Majorana mass terms.
\item What is the absolute mass scale (with implications for cosmology)? This is very difficult,
but ordinary and double beta decay experiments, as well as future CMB experiments, may
be able to establish the scale.
\item Is the hierarchy ordinary or inverted? Possibilities include matter effects in future long-baseline
experiments or in the observation of a future supernova, the observation of $\beta \beta_{0\nu}$ if the neutrinos are Majorana, and possibly effects in future Solar neutrino experiments.
\item What is $\theta_{13}$? This is especially important because the observation of leptonic CP violation requires a non-zero $\theta_{13}$. This will be addressed in a program of future reactor and long-baseline experiments, and possibly at a dedicated neutrino factory (from a muon storage ring).
\item Why are the mixings large, while those of the quarks are small? For example, the simplest
grand unified theory seesaws would lead to comparable mixings, although this can be evaded in
more complicated constructions.
\item If the LSND result is confirmed, it will suggest mixing between ordinary and sterile neutrinos, presenting
a serious challenge both to particle physics and cosmology, or imply something even more bizarre, such
as CPT violation.
\item Are there any new $\nu$ interactions or anomalous properties such as large magnetic moments? Most such ideas are excluded as the dominant effect for the Solar and atmospheric neutrinos, but
could still appear as subleading effects.
\end{itemize}
Answering these questions and unraveling the origin of the masses is  therefore an important 
and exciting probe of  new particle physics. Future Solar neutrino experiments, observations of
the neutrinos from a future core-collapse supernova, and observation of high energy neutrinos in
large underground/underwater experiments would also be significant probes not only of the
neutrino properties but also the underlying astrophysics. 

\section*{Acknowledgements}

This work was supported in part by CONACYT (M\'exico) contract 42026--F, 
by DGAPA--UNAM contract PAPIIT IN112902, and by the
U.S.\ Department of Energy under Grant No.\ DOE-EY-76-02-3071.


\end{document}